# Phonon Raman scattering of $R$CrO$_3$ perovskites ($R$ = Y, La, Pr, Sm, Gd, Dy, Ho, Yb, Lu)

M. Weber[1], J. Kreisel[1,2,*], P.A. Thomas[1], M. Newton[1], K. Sardar[3], R.I Walton[3]

[1] Department of Physics, University of Warwick, Coventry CV4 7AL, United Kingdom

[2] Laboratoire Matériaux et Génie Physique, Minatec, CNRS, Grenoble Institute of Technology, 3, parvis Louis Néel, 38016 Grenoble, France

[3] Department of Chemistry, University of Warwick, Coventry CV4 7AL, United Kingdom

**Abstract**

We report a systematic investigation of orthorhombic perovskite-type $R$CrO$_3$ powder samples by Raman scattering for nine different rare earth $R^{3+}$ cations ($R$ = Y, La, Pr, Sm, Gd, Dy, Ho, Yb, Lu). The room-temperature Raman spectra and the associated phonon mode assignment provide reference data for structural investigation of the whole series of $R$CrO$_3$ orthochromites and phonon ab-initio calculations. The assignment of the chromite spectra and comparison with Raman data on other orthorhombic perovskites allows correlating the phonon modes with the structural distortions in the $R$CrO$_3$ series. In particular, two $A_g$ modes are identified as octahedra rotation soft modes as their positions scale linearly with the octahedra tilt angle of the CrO$_6$ octahedra.

[*] to whom correspondence should be addressed (E-mail: Jens.Kreisel@grenoble-inp.fr)

# I. INTRODUCTION

The understanding of functional $ABO_3$ perovskite-type oxides is a very active research area with relevance to both fundamental- and application-related issues. A particular aspect of perovskites is its capacity to adopt a multitude of different structural distortions due to the possible incorporation of almost every element of the periodic table into its structure [1]. Such distortions can be driven by external parameters like temperature, pressure or chemical composition, which leads to an extraordinary richness of physical properties within the family of perovskites. Multiferroic perovskites, which possess simultaneously several so-called ferroic orders such as ferromagnetism, ferroelectricity and/or ferroelasticity, currently attract a specific interest [2-4]. Structural distortions in perovskites can be described by separating three main features with respect to its ideal cubic structure [1, 5-6]: (*i*) a rotation (tilt) of essentially rigid $BO_6$ octahedra, (*ii*) polar cation displacements which often lead to ferroelectricity (*iii*) distortions of the octahedra such as the Jahn-Teller distortion. Past investigations of these instabilities have been a rich source for the understanding of structural properties not only in perovskites but also in oxides in general. The most common distortion in perovskites is the tilt of octahedra. Although many different octahedral tilt systems exist [1, 5-7], the $R\text{-}3c$ structure with tilt system $a^-a^-a^-$ (Glazer's notation [6]) and the $Pnma$ with $a^-b^+a^-$ are by far the most common structures found in perovskites.

We are particularly interested in perovskites with the orthorhombic $Pnma$ structure which is for instance adopted by $R$FeO$_3$ orthoferrites, $R$MnO$_3$ orthomanganites, $R$NiO$_3$ orthonickelates or $R$ScO$_3$ orthoscandates ($R$ = Rare earth). In all these perovskites the orthorhombic distortion (octahedral tilt angle) can be continuously tuned by the ionic size $r_R$ of the $R^{3+}$ rare earth. This rotation is equivalent to modify the $B$-O-$B$ bond angle and the $B$-O orbital overlap, which in turn leads to a rich magnetic phase diagram for magnetic $B$-cations, accompanied by intriguing metal insulator phase transitions when $B$ = Mn, Ni. In this study we focus on the family of $R$CrO$_3$ orthochromites, which have in the past attracted considerable interest for their complex magnetic phases at low temperature [8-11] and, more recently, for their potential magnetoelectric or multiferroic properties [11-15], similarly to manganites and nickelates.

The aim of our study is to provide a better understanding of the phonon spectra of $R$CrO$_3$ and to correlate individual phonon modes to the structural distortions of the structure. For this we have investigated $R$CrO$_3$ powder samples with a large number of different rare earth (Y, La, Pr, Sm, Gd, Dy, Ho, Yb, Lu) by phonon Raman spectroscopy, which is known to be a versatile technique for the investigation of both phase transitions [16-23] and more subtle structural distortions in perovskites [24-29]. While the Raman signature of orthomanganites [24-27, 30-31] or orthoferrites [32-34] is well known and understood, investigations of orthochromites remain limited to Raman studies for $R$ = Y, La, Nd, Gd, Ho, Er [22, 24, 35-38], out of which the single crystal work on LaCrO$_3$ [22] and YCrO$_3$ [24] is of particular relevance for this work. However, the whole $R$CrO$_3$ series has not yet been systematically investigated.

## II. EXPERIMENTAL

Powder samples of $R$CrO$_3$ ($R$ = Y, La, Pr, Sm, Gd, Dy, Ho, Yb, Lu) chromites were synthesized by a direct hydrothermal synthesis that gave polycrystalline powders of a greenish colour. More details about the synthesis of samples can be found in ref. [10]. The high-quality of the samples has been attested by previous X-ray diffraction measurements and the characterization of their magnetic properties [10]. Throughout this manuscript we will discuss the different structural details of orthochromites, Table 1 presents a resume of these characteristics for both the chromites investigated here and others discussed in the text. A particular attention has been paid to the calculation of the octahedral tilt angles. This is not straightforward for the *Pnma* structure as the angles can be calculated either from cell parameters or from the atomic coordinates [1], both results are compared in Table 1. Typically, the use of cell parameters introduces an error in the tilt angles by about 15 % as the octahedra are slightly distorted and we thus rather rely on the tilt angles which we have calculated from the atomic coordinates according to the formalism in ref. [39]. In order to illustrate the distortion throughout the series, Figure 1 presents the evolution of the lattice parameters and the tolerance factor $t$, defined as $t = (r_{R3+} + r_{O2-}) / \sqrt{2} (r_{Cr3+} + r_{O2-})$.

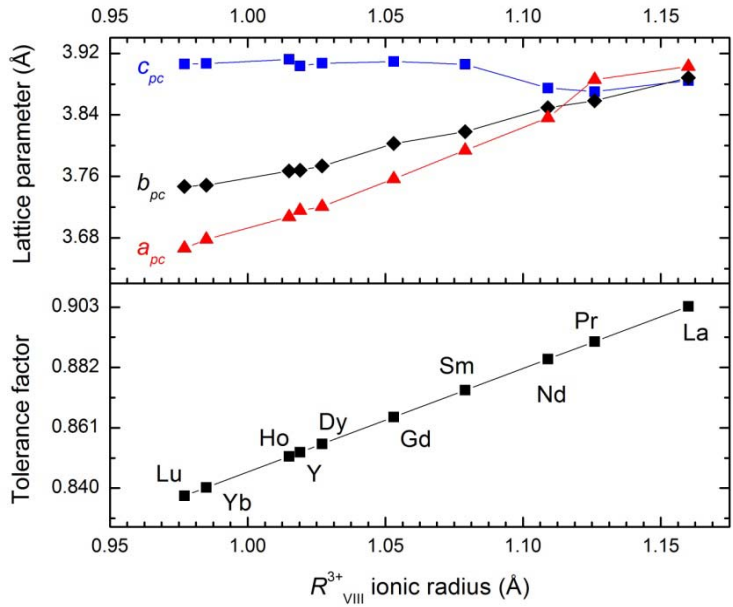

**Figure 1**
Variation of the pseudo-cubic cell parameters and the tolerance factor for $R$CrO$_3$ as a function of the $R^{3+}_{VIII}$ ionic radius. The lattice parameters for NdCrO$_3$ are taken from ref. [38], for all other chromites from ref. [10].

| | Lattice parameters | | | | | | CrO$_6$ octahedra tilt angles | | | |
| | | | | | | | From lattice parameters | | From atomic positions | |
| | r$_{R3+}$ (Å) | a (Å) | b (Å) | c (Å) | V (Å$^3$) | t | φ$_{[010]}$ (°) | θ$_{[101]}$ (°) | φ$_{[010]}$ (°) | θ$_{[101]}$ (°) |
|---|---|---|---|---|---|---|---|---|---|---|
| LaCrO$_3$ | 1.160 | 5.494 | 7.777 | 5.520 | 235.814 | 0.903 | 2.50 | 5.57 | 5.25 | 10.92 |
| PrCrO$_3$ | 1.126 | 5.473 | 7.717 | 5.496 | 232.104 | 0.891 | | 5.20 | | |
| NdCrO$_3$ | 1.109 | 5.480 | 7.699 | 5.425 | 228.8 | 0.885 | 4.79 | 8.12 | 7.55 | 13.86 |
| SmCrO$_3$ | 1.079 | 5.524 | 7.637 | 5.366 | 226.328 | 0.874 | 6.45 | 13.73 | | |
| GdCrO$_3$ | 1.053 | 5.529 | 7.605 | 5.313 | 223.395 | 0.864 | 8.91 | 16.06 | 10.96 | 14.39 |
| DyCrO$_3$ | 1.027 | 5.526 | 7.547 | 5.262 | 219.428 | 0.855 | 9.61 | 17.77 | 12.09 | 17.19 |
| YCrO$_3$ | 1.019 | 5.520 | 7.536 | 5.255 | 218.606 | 0.852 | 9.56 | 17.85 | 12.20 | 16.98 |
| HoCrO$_3$ | 1.015 | 5.533 | 7.534 | 5.243 | 218.554 | 0.851 | 10.21 | 18.62 | | |
| ErCrO$_3$ | 1.004 | 5.503 | 7.504 | 5.214 | 215.303 | 0.847 | 10.70 | 18.66 | 12.23 | 17.32 |
| YbCrO$_3$ | 0.985 | 5.525 | 7.497 | 5.202 | 215.462 | 0.840 | 11.13 | 19.70 | | |
| LuCrO$_3$ | 0.977 | 5.524 | 7.494 | 5.186 | 214.667 | 0.837 | 11.89 | 20.16 | | |

**Table 1:**
Structural characteristics of $R$CrO$_3$ samples: lattice parameters in the orthorhombic setting, $R^{3+}$ ionic radii ($r_{R3+}$ values given in an eight-fold environment from [40]) and octahedral tilt angles (φ[010], θ[101]). The tilt angles are determined both from lattice parameters [1] and, where available, from atomic coordinates [39]. The lattice parameters for NdCrO$_3$ are taken from ref. [38], for ErCrO$_3$ from [41] and for all other chromites from ref. [10]. The atomic positions for the calculation of tilt angles are taken from refs. [41-45].

Raman spectra were recorded with a Renishaw inVia Reflex Raman Microscope with a low-wavenumber spectral cutoff at about 120 cm$^{-1}$. Experiments were conducted in micro-Raman mode at room temperature by using 633 nm He-Ne laser as exciting wavelength. It is well-known that Raman spectra recorded on transition-metal oxides such as orthomanganites [30, 46] or orthonickelates [29, 47] often show a strong dependence on the exciting laser power leading to structural modifications, phase transitions or even locally decomposed material. Contrary to the above samples, our $R$CrO$_3$ powders are not black where laser heating is naturally expected. Nevertheless, we have observed that experiments with a laser power above 10 mW lead to a significantly modified spectral signature $R$CrO$_3$ in terms of band position, width and intensity due to local laser heating. As a consequence, our experiments have been carried out using laser powers of less than 1 mW under the microscope, and we have carefully verified that no structural transformations or overheating take place.

## III. RESULTS AND DISCUSSION

The ideal cubic structure of $AB$O$_3$ perovskites is rather simple, with corner-linked anion octahedra $B$O$_6$, the $B$ cations at the centre of the octahedra and the $A$ cations in the space between the octahedra. In this cubic perovskite structure Raman scattering is forbidden by symmetry.

$R$CrO$_3$ chromites crystallize in an orthorhombically distorted perovskite structure with space group $Pnma$ (alternative setting $Pbnm$). With respect to the ideal cubic $Pm\bar{3}m$ perovskite structure this orthorhombic *Pnma* structure is obtained by an anti-phase tilt of the adjacent CrO$_6$ octahedra ($a^-b^+a^-$ in Glazer's notation [6]). The tilting of the CrO$_6$ octahedra necessarily induces a distortion of the $R$O$_{12}$ polyhedra. In the *Pnma* structure, the *A*-cation is usually considered to be in eight-coordination and also antiparallel *A*-cation displacements are permitted by symmetry. Cr$^{3+}$ is a Jahn-Teller (JT) in-active cation, leading to almost no dispersion in the Cr-O bonds, typically an order of magnitude lower than what is observed in JT-active orthomanganites. All three distortions (tilt, octahedra distortion and *A*-cation displacement) break the cubic symmetry and thus activate zone centre Raman modes. According to group theory, the orthorhombic *Pnma* structure with four formula units per unit cell give rise to 24 Raman-active modes [30]: 7 $A_g$ + 5 $B_{1g}$ + 7 $B_{2g}$ + 5 $B_{3g}$.

Raman spectra of orthochromites with $R$ = La, Pr, Sm, Gd , Dy , Ho ,Yb, Lu are presented in Figure 2 and show 11 to 14 modes depending on the rare earth. The other predicted modes are either of too low intensity to be observed or below our experimental cut-off. Our powder Raman spectra of LaCrO$_3$, GdCrO$_3$ and HoCrO$_3$ are consistent with literature work on single crystals of the same materials [22, 24, 35-36], for the other five chromites there is to date no data in the literature, to the best of our knowledge. Table 2 presents the position and the hereafter discussed assignment of the individual modes, together with their notation used in the manuscript.

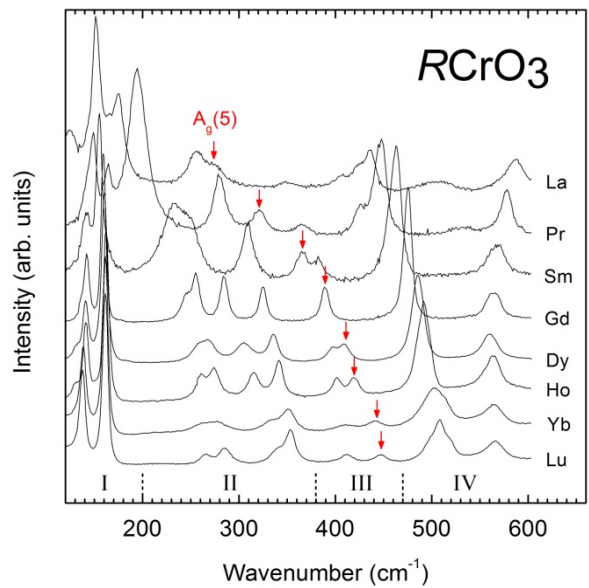

**Figure 2:**
Raman spectra at 300K of $R$CrO$_3$ powders ($R$ = *La, Pr, Sm, Gd , Dy, Ho ,Yb, Lu*) collected with a 632.8 nm He-Ne laser line. I, II, III, and IV denote specific spectral regions discussed in the text. The red arrows illustrate for a specific mode, how the phonons can be followed though the series of $R$CrO$_3$.

Qualitatively, the overall spectral signature of the chromites is similar, in agreement with their shared space group where the orthorhombic distortion varies continuously from the smallest rare earth Lu ($r_{Lu,VIII}$ = 0.98 Å) to the largest rare earth La ($r_{La,VIII}$ = 1.16 Å) [10, 40]. According to Figure 1, LaCrO$_3$ and PrCrO$_3$ occupy a special place in the orthochromite series in the sense that their orthorhombic distortion is significantly smaller than that of other orthochomites with very close lattice parameters *a*, *b* and *c*.

Consistently, the spectra of LaCrO$_3$ and PrCrO$_3$ differ the most from the other spectra of the series, namely in the low- and mid-wavenumber range, suggesting that the Raman signature can be used to follow the orthorhombic distortion.

**Table 2**
Band positions and assignment of the observed Raman modes in $R$CrO$_3$. While the symmetry assignment is discussed in text, the activating distortions and main atomic motions are extended from refs. [22, 24, 26].

| Symmetry | LuCrO$_3$ | YbCrO$_3$ | HoCrO$_3$ | YCrO$_3$ | DyCrO$_3$ | GdCrO$_3$ | SmCrO$_3$ | PrCrO$_3$ | LaCrO$_3$ | Activating distortion | Main atomic motions |
|---|---|---|---|---|---|---|---|---|---|---|---|
| A$_g$(1) | | | | 149.0 | | | | | | | |
| A$_g$(2) | 137.6 | 138.5 | 141.2 | 184.5 | 141.0 | 142.2 | 140.8 | 143.5 | 175.2 | rot[101] | A(z) out of plane |
| A$_g$(3) | 285.9 | 278.0 | 274.3 | 281.8 | 270.2 | 255.5 | 231.5 | 193.2 | 104.0 | rot[010], JT | BO$_6$ in-phase y rotations |
| A$_g$(4) | 353.4 | 351.7 | 341.3 | 338.8 | 335.6 | 324.9 | 309.1 | 279.9 | 255.3 | A shift | O1(x), A(-x) |
| A$_g$(5) | 446.8 | 441.7 | 419.6 | 421.1 | 409.5 | 392.4 | 365.7 | 320.9 | 274.7 | rot[101] | BO$_6$ out-of-phase x rotations |
| A$_g$(6) | 508.2 | 501.9 | 493.3 | 489.1 | 486.1 | 477.0 | 465.3 | 449.7 | 436.6 | rot[101] | BO$_6$ bendings |
| A$_g$(7) | | | | 558.3 | | 561.2 | | | | | |
| B$_{1g}$(1) | 265.6 | 260.7 | 260.3 | 273.1 | 260.5 | 245.3 | 244.7 | - | - | | |
| B$_{1g}$(2) | 411.8 | 408.8 | 401.6 | 405.4 | 397.2 | 388.7 | 382.4 | 365.4 | 351.4 | A shift | A(z), O1(-z) |
| B$_{2g}$(1) | 161.3 | 161.4 | 162.1 | 216.8 | 160.4 | 159.3 | 155.2 | 149.1 | 151.7 | rot[101] | A(x) |
| B$_{2g}$(2) | 339.8 | 332.4 | 315.2 | 316.7 | 305.5 | 284.5 | 252.7 | 201.1 | 124.6 | A shift | A(z), O1(z) |
| B$_{2g}$(3) | 519.2 | 514.1 | 493.3 | 498.3 | 490.6 | 473.6 | 460.9 | 425.5 | 405.4 | rot[101] | BO$_6$ out-of-phase bendings |
| B$_{3g}$(1) | | | 129.2 | 171.5 | 130.5 | 137.1 | | | | rot[101] | A(y) |
| B$_{3g}$(2) | 497.4 | 504.7 | 486.9 | 482.3 | 482.2 | 473.6 | 454.7 | 443.8 | 424.5 | rot[101] | out-of-phase O2 "scissors"-like |
| B$_{3g}$(3) | 565.1 | 563.8 | 562.9 | 540.5 | 560.0 | 568.7 | 567.2 | 577.7 | 586.8 | rot[101] | O2,O1 antistretching |

## A. Symmetry assignment of Raman modes in $R$CrO$_3$

A symmetry assignment of the different Raman features is the *sine qua non* condition for a later discussion of the correlation between structural distortions and the Raman signature. Generally speaking, the use of powder samples with a random orientation of crystallites, as in our study, inhibits the assignment of Raman modes by using polarized configurations. As a consequence, the hereafter presented assignment is in a first step based on single crystal measurements that have been reported in the literature for LaCrO$_3$ [22] GdCrO$_3$ [35], YCrO$_3$ [24, 35] and HoCrO$_3$ [36]. In a second step, we have taken advantage of the continuous changes in the Raman spectra across the whole $R$CrO$_3$ series, which has allowed verifying the reported literature assignments, assigning the modes for the not yet reported samples and investigating the crossing of several modes. For the sake of clarity and given the important number of both materials and Raman bands, we will successively discuss four spectral regions as defined in Figure 2:

- *Region I* (below 200 cm$^{-1}$) characterized by two sharp and intense bands.
- *Region II* (200 to 380 cm$^{-1}$) characterized by four partly overlapping bands.
- *Region III* (380 to 500 cm$^{-1}$) with two bands of similar intensity, except for GdCrO$_3$ where only one band is observed.
- *Region IV* (500 to 600 cm$^{-1}$) with two features, each characterized by overlapping bands.

### 1. *Region I (100 to 200 cm$^{-1}$)*

In the approximation of an harmonic oscillator $\omega = (k/\mu)^{1/2}$ ($k$, force constant; $\mu$, reduced mass) the heaviest atom of the structure is expected to vibrate in the low-wavenumber region. Among the materials shown in Figure 2 the mass increases only by about 25 % from La to Yb, so that the frequency associated with *A*-cation vibrations is expected to increase only little with decreasing mass. Bands which are dominated by vibrations of the *A*-cation are expected to be only little affected by changes in the orthorhombic distortion.

The two sharp low-wavenumber bands at 137 and 161 cm$^{-1}$ (values for LuCrO$_3$) show only small variation despite the large changes in distortion throughout the series. It is thus natural to assign these two bands to *A*-cation vibrations.

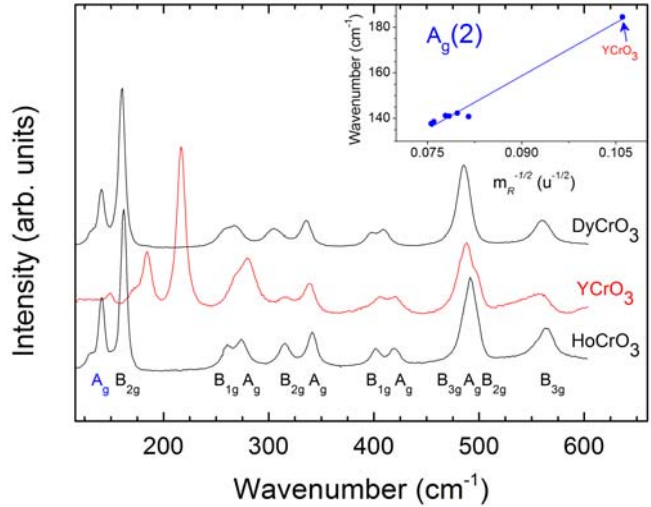

**FIG. 3:**
Comparison of Raman spectra for of HoCrO$_3$ (bottom, black), YCrO$_3$ (middle, red) and DyCrO$_3$ (top, black) illustrating a significant shift of the low wavenumber modes for YCrO$_3$. The assignment of $A_g$, $B_{1g}$, $B_{2g}$ and $B_{3g}$ modes is based on ref. [24] for YCrO$_3$. The inset shows the mass-dependent evolution of the two low-wavenumber bands.

The most convincing argument for the assignment of the two low-wavenumber bands to *A*-cation vibrations comes from Figure 3 which compares the Raman spectrum of YCrO$_3$ to that of DyCrO$_3$ and HoCrO$_3$. The three rare earth have close ionic radii for Dy ($r_{Dy,VIII}$ = 1.03 Å), Y ($r_{Y,VIII}$ = 1.02 Å) and Ho ($r_{Ho,VIII}$ = 1.02 Å) leading to a very similar orthorhombic distortion, but the mass of the Y ($m_Y$ = 89) is very different to that of Dy ($m_{Dy}$ = 162.5) and Ho ($m_{Ho}$ = 165). It can be seen from Figure 3 that the sharp bands shift for YCrO$_3$ significantly by about 45 cm$^{-1}$, while the other bands are only little or not affected. The inset of Figure 3 shows that this shift corresponds for the band at 137 cm$^{-1}$ well to the mass-induced shift expected from a harmonic oscillator. This observation provides conclusive evidence that the two sharp low-wavenumber modes are largely dominated by vibrations of the *A*-cation. Following the symmetry assignment from single crystal work on YCrO$_3$ [24], we assign the lower mode to $A_g$ and the higher mode to $B_{2g}$ symmetry. Beyond clarifying the band assignment it is interesting to note that Figure 3 illustrates that materials like YCrO$_3$ or HoCrO$_3$, which present an almost identical distortion and thus a rather similar XRD pattern, can be easily differentiated by Raman scattering.

### 2. *Region II (200 to 380 cm$^{-1}$)*

*Region II* is characterized by two doublets, which can be assigned by extension of single crystal literature data [24, 35-36] to bands of $B_{1g}$ symmetry at 265 cm$^{-1}$, $A_g$ at 285 cm$^{-1}$, $B_{2g}$ at 339 cm$^{-1}$ and $A_g$ at 353 cm$^{-1}$ (values for LuCrO$_3$). Different spectral features should be commented: (*i*) Figure 3 shows that the $B_{1g}$-$A_g$ doublet around 275 cm$^{-1}$ is affected by changes in the mass of the *A*-cation, while the doublet around 340 cm$^{-1}$ is not. (*ii*) The $B_{2g}$ mode shows a very significant low-wavenumber shift with increasing radii of the rare earth (decreasing distortion), making its attribution difficult for LaCrO$_3$ and PrCrO$_3$. (*iii*) As an overall trend, we note that this spectral region is greatly affected by changes in the orthorhombic distortion, which becomes namely obvious for the materials with large $r_R$ (thus reduced orthorhombicity), as will be discussed later.

### 3. *Region III (380 to 500 cm$^{-1}$)*

*Region III* is characterized by two bands of similar intensity, the evolution of their band position across the *R*CrO$_3$ series is presented in more detail in Figure 4. For the smallest $R^{3+}$ in LuCrO$_3$ the two bands at 411 and 446 cm$^{-1}$ are well separated and can be assigned to $B_{1g}$ and $A_g$ symmetry, respectively. Due to a pronounced low-wavenumber-shift of the $A_g$(5) band with increasing $r_R$, the $B_{1g}$(2) and $A_g$(5) modes gradually approach until only a single and more intense band is observed for GdCrO$_3$. Upon further increase of $r_R$ the two modes gradually split again. This continuous evolution of both bands and the observation of a single intense band for GdCrO$_3$ strongly suggest a mode-crossing. As the two modes have different $B_{1g}$ and $A_g$ symmetry such a mode crossing is allowed by symmetry. The low-wavenumber-shift of the $A_g$(5) band throughout the whole series of nine rare earth is remarkable and the most pronounced shift observed in the spectra. For the case of LaCrO$_3$ we note that the $A_g$(5) band has shifted to such an extent that it closely approaches a lower lying $A_g$(4) mode of *region II* (see Figure 2). Such a closeness of modes with the same symmetry can result in a mixing of the modes, as reported in the similar orthomanganites [26], although we note that the intensity of the modes suggests that a transfer of the character has not yet been occurred. The origin of the pronounced shift of the $A_g$(5) band will be discussed later in more detail.

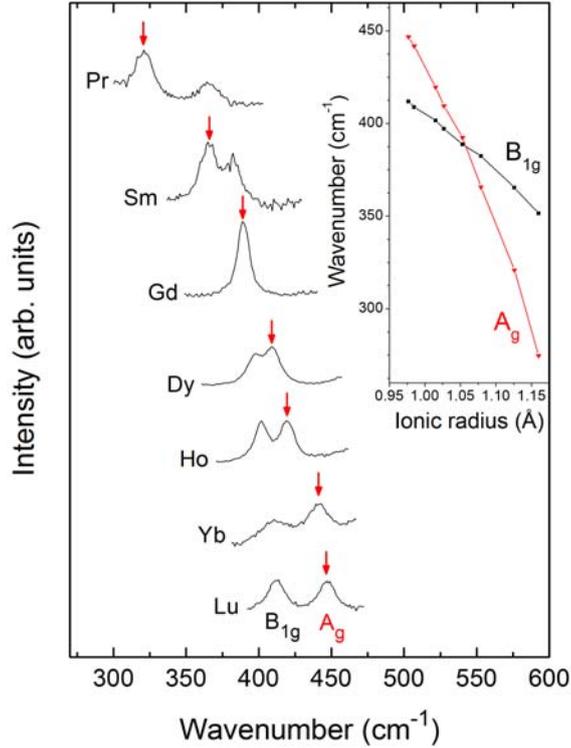

**Figure 4:**
Detailed view of the mid-wavenumber spectral region III. (a) Raman modes and (b) Evolution of the band positions with $r_R$ for $R$CrO$_3$.

### 4. Region IV (500 to 600 cm$^{-1}$)

*Region IV* looks at first sight rather simple with only two broad spectral features, but a closer inspection of Figure 2 shows that it is in fact a rather complex region with an important number of overlapping and crossing bands. The presence of three overlapping features at 497, 508 and 519 cm$^{-1}$ is best observed for the most distorted LuCrO$_3$ and can be assigned to $B_{3g}$, $A_g$ and $B_{2g}$ symmetry, respectively. The different symmetries allow these three modes to cross or take the same frequency as for instance for GdCrO$_3$ where only a single intense band is observed around 475 cm$^{-1}$. Due to the closeness and intercrossing of the three bands, a band assignment of the individual shoulders is only possible on the basis of polarized Raman data on single crystals (available in literature for some rare earths [22, 24, 35-36]) and on the hypothesis that the $A_g$ mode remains throughout the series the most intense among the three modes. Based on this, we observe that the $B_{2g}$ mode shows the most pronounced low-wavenumber shift and even crosses the two other modes from the high-wavenumber-side in LuCrO$_3$ to the low-wavenumber-side of the three-mode feature in LaCrO$_3$.

The broadness of the massif at around 580 cm$^{-1}$ suggests two or three underlying bands for this feature, which can unfortunately not be distinguished by our powder Raman experiment, although we note that their presence is suggested by single crystal data in the literature [22, 24, 35-36]. Modes in this region involve mainly stretching vibrations of the $CrO_6$ octahedra, which in turn explains their insensitivity to the orthorhombic distortion, similar to observations for equivalent modes in manganites [26].

## 5. The particular case of LaCrO$_3$, PrCrO$_3$ and SmCrO$_3$

Figure 2 and the discussion in the previous sections show that the Raman spectra of LaCrO$_3$, PrCrO$_3$ and SmCrO$_3$ present in the low-wavenumber region a spectral signature that is significantly different from the other $R$CrO$_3$. This difference, which is in agreement with the reduced distortion of these materials when compared to the other rare earth (see Figure 1 and Table 1), complicates the assignment of the individual modes as the spectral changes are no more smooth. Figure 5 presents a zoom into the 100 to 450 cm$^{-1}$ region for LaCrO$_3$, PrCrO$_3$ and SmCrO$_3$, which illustrates that the assignment of the 300 to 400 cm$^{-1}$ region is still straightforward. However, the region below 300 cm$^{-1}$ is more complex with two $A_g$ modes and two $B_{2g}$ modes and where modes of the same symmetry are likely to interact and mix.

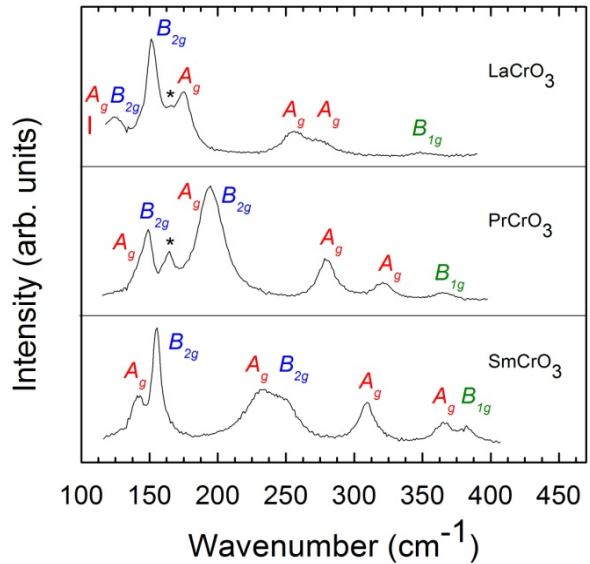

**Figure 5:**
Detailed view and assignment of the Raman modes in the low-wavenumber and mid-wavenumber region for LaCrO$_3$, PrCrO$_3$ and SmCrO$_3$. The band at 160 cm$^{-1}$, marked with * is a spurious line, likely related to an impurity phase. The assignment of LaCrO$_3$, specifically the position and assignment of the 104 cm$^{-1}$ $A_g$ band marked by a red line, is from ref. [22].

We first discuss the behavior of the $B_{2g}$ modes. It is tempting to relate the lower $B_{2g}$ modes to one type of vibrational mode and the higher lying $B_{2g}$ to a second mode. However, Figure 5 shows that the lower $B_{2g}$ mode in LaCrO$_3$ is weak, which is in contrast to its intense and sharp profile throughout the whole series of $R$CrO$_3$ (including SmCrO$_3$) and the higher $B_{2g}$ mode of LaCrO$_3$. This observation suggests a transfer of intensity which is the classic behavior for mixing of modes of the same symmetry and which are close in wavenumber, resulting in transfer of their vibrational character. To support this scenario we consider the mode assignments of lattice-dynamical calculations (LDC) reported for LaCrO$_3$ [22] and YCrO$_3$ [24] which lead to the following scenario: For all $R$CrO$_3$ (except LaCrO$_3$ and PrCrO$_3$) the higher $B_{2g}(2)$ mode is activated by the $A$-cation shift [24] while the lower $B_{2g}(1)$ mode is activated by octahedral rotations around [101]. For PrCrO$_3$ we expect that the two modes mix and transfer energy although the overlap of the modes inhibits a detailed analysis. Finally, for LaCrO$_3$ the modes have clearly exchanged their character as evidenced by the intensity transfer, but our data does not allow judging to what extend the modes are still partly mixed. As a consequence, we assign the 124 cm$^{-1}$ mode in LaCrO$_3$ to $B_{2g}(2)$ and the 154 cm$^{-1}$ mode to $B_{2g}(1)$.

The discussion of the $A_g$ modes on the basis of our own measurements is more difficult as their difference in terms of intensity is less marked and namely because our spectral cut-off does not allow us the observation of the low lying $A_g$ mode of LaCrO$_3$, which was reported to be located at 104 cm$^{-1}$ [22]. Our discussion thus has to rely on the mode assignments of lattice-dynamical calculations (LDC) reported for LaCrO$_3$ [22] and YCrO$_3$ [24]. According to [24], the higher $A_g(3)$ mode of $R$CrO$_3$ (except LaCrO$_3$ and PrCrO$_3$) has the pattern of octahedra rotation, while the lower $A_g(2)$ is dominated by $A$-cation vibrations. From our data with overlapping modes it is difficult to judge if the two modes mix for PrCrO$_3$. However, according to the assignment presented in ref. [22] the two modes have clearly exchanged their character for LaCrO$_3$ so that we assign the 104 cm$^{-1}$ mode in LaCrO$_3$ to $A_g(3)$ and the 175 cm$^{-1}$ mode to $A_g(2)$.

As a consequence, the low wavenumber region of LaCrO$_3$, PrCrO$_3$ and SmCrO$_3$ is characterized by a complex crossing in-between two $A_g$ modes and in-between two $B_{2g}$ modes. Further understanding of the mode mixing could be gained from investigations of solid solutions such as La$_{1-x}$Pr$_x$CrO$_3$ or La$_{1-x}$Sm$_x$CrO$_3$.

## B. Phonon Raman modes *vs*. ionic radii & structural distortions

Figure 6 presents the evolution of the band position for the different chromites as a function of the rare earth $R^{3+}$ ionic radii. The proposed assignment in terms of symmetry and mode crossing is based on the above discussion of the individual spectral regions. The figure illustrates the overall trend of decreasing band positions with increasing $r_R$ which naturally correlates with the increase of most bond-lengths. The few exceptions to this general trend have been discussed in the previous section.

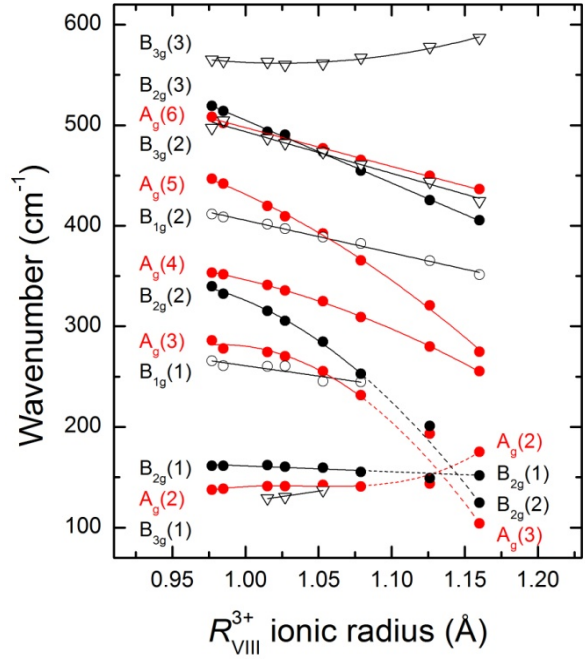

**Figure 6:**
Raman phonon wavenumbers of $R$CrO$_3$ as a function of the rare earth $R^{3+}$ ionic radius. All lines are guides to the eye only. The dashed lines for the low-wavenumber modes indicate schematically the region of mode mixing, discussed in the text.

It can be seen that the $r_R$-dependent shift in wavenumber varies significantly among the different Raman modes and it is natural to expect that the modes with a significant shift involve O-Cr-O bending (thus octahedra tilting), as this is the main structural distortion that changes throughout the series. Before discussing this in more detail, we shall recall early work by Scott and co-workers [16-17, 19] which demonstrated that Raman spectroscopy allows investigating the distortion of perovskites that is caused by the rotation of octahedra. By studying temperature-dependent Raman spectra of SrTiO$_3$ [16] and LaAlO$_3$ [17] across their phase transition from a non-cubic structure with octahedra tilts, to the undistorted cubic structure, they have observed that the phase transition is driven by soft modes which scale with the angle of rotation of the octahedra. This Raman soft-mode picture has later shown to describe also the behavior of the same model materials under high-pressure [20, 23, 48-49]. As this soft mode concept is of importance for the interpretation of our data, we will shortly illustrate its application to the relatively simple model system LaAlO$_3$ (LAO). LAO crystallizes in a rhombohedral $R$-$3c$ structure that is characterized by an anti-phase $a^-a^-a^-$ tilt of the adjacent AlO$_6$ octahedra about the [111]$_p$ pseudo-cubic diagonal which gives rise to five Raman-active modes $\Gamma_{Raman} = A_{1g} + 4\,E_g$. The $A_{1g}$ and one $E_g$ mode of the rhombohedral structure are mainly related to the collective modes of the octahedral network. The vibration of the $A_{1g}$ mode has specifically the pattern of the single rotation of adjacent AlO$_6$ about the [111] axis [17, 27, 50-51]. Experimentally, the position of the $A_{1g}$ modes shows a linear scaling behavior with the AlO$_6$ octahedral rotation angle with a slope of about 23.5 cm$^{-1}$/degree [50-51]. LaNiO$_3$, which adopts the same structure but has a different chemistry presents interestingly a similar slope of 23 cm$^{-1}$/degree [47, 52]. The observed linear relationship is in agreement with Landau theory which predicts for the simplest second-order phase transition that the soft-mode frequency $\omega$ should vary in the distorted phase ($T < T_c$) with temperature as $\omega^2 = A(T_c-T) \sim \theta^2$, where $\theta$ is the octahedral rotational order parameter [53].

The situation is more complicated in perovskites with a *Pnma* structure, because several soft modes take part in the lattice distortion. The structure can either be described in terms of three orthogonal tilt angles tilt angles (leading to the $a^-b^+a^-$ Glazer's notation) or by rotations $\theta$, $\phi$ and $\Phi$ around the pseudo-cubic $[101]_{pc}$, $[010]_{pc}$, and $[111]_{pc}$ axes, respectively [1]. Under the assumption that the orthogonal rotations $a^-_x$ and $a^-_z$ are approximately equal in the *Pnma* structure, it has been shown by Megaw [54] that the tilt angles $\theta$ and $\phi$ alone are sufficient for describing the octahedral tilting of the structure while the angle $\Phi$ can be obtained via cos $\Phi$ = cos$\theta$ cos$\phi$ [1]. The necessary requirement for correlating the two tilt angles $\theta$ and $\phi$ to the observed Raman signature is the knowledge of the distortion-dependent soft modes, for which Iliev *et al.* [26] have shown that they are of $A_g$ symmetry. This identification of this mode is not trivial, given the important number of Raman-active modes in the *Pnma* structure. Also the *Pnma* structure is among the most stable structures within perovskites, so that usually only a weak softening is observed, complicating the identification of the vibrational modes which present the pattern of the distortion.

In the past, the empirical comparison of several materials with the same type of structural distortion has been used to overcome this problem and has been a rich source of understanding Raman signatures in oxide materials. For the specific case of orthorhombic perovskites we mention the systematic Raman work on orthorhombic *R*MnO$_3$ orthomanganites [26, 30-31] and *R*ScO$_3$ orthoscandates [55-57], all of which crystallize in the same *Pnma* symmetry as orthochromites. These studies have allowed identifying two $A_g$ modes with a octahedra-rotation vibrational pattern of which the frequency scale linearly with the rotation angle of the $B$O$_6$ octahedral tilting [26, 55]. Interestingly the observed slopes, 23.5 cm$^{-1}$/degree for manganites [26] and 20 cm$^{-1}$/degree for scandates [55] are close to the value observed for rhombohedral perovskites (see above).

We now discuss the observed spectral changes in our orthochromites. The inspection of Figure 6 shows that the $A_g$(5) mode presents a particularly pronounced shift in position with increasing $r_A$, it is thus natural to assign this mode to one of the two $A_g$ octahedral soft modes which are expected to scale with one of the octahedra tilt angles. Figure 7 plots the $A_g$(5) phonon positions as a function of octahedra tilt angles for those $R$CrO$_3$ where available atomic coordinates allow a reliable calculation of the octahedra tilt angles (see tables 1 and 2 for explicit values). It can be seen that the positions of the $A_g$(5) mode in $R$CrO$_3$ shows the same tendency as the manganite and scandate modes, i.e. they scale linearly with the octahedra tilt angles at a slope of $\approx$ 24.3 cm$^{-1}$/degree, adding further support to the assignment of the $A_g$(5) mode. The visual assignment of the second soft mode is less straightforward as no other $A_g$ Raman mode shifts for the small *A*-cations in a similarly strong way as $A_g$(5). However, the earlier work on manganites and scandates suggests that both $A_g$ soft modes lie on the same slope (see [26, 55] and Figure 7). As a consequence of this, we have plotted the evolution of all $A_g$ modes against the second tilt mode by assuming the 24.3 cm$^{-1}$/degree slope observed for $A_g$ (5). This plot (not shown) illustrates that only the evolution of the $A_g$ (3) mode follows the behavior of the $A_g$ (5), allowing assigning the $A_g$ (3) mode as the second soft mode, in agreement with earlier assignments of LaCrO$_3$ [22] and YCrO$_3$ [24] As a matter of fact, this identification has also provided confidence to our assignment of the low-frequency region (see section III.A.5).

On the basis of these findings, we conclude that the $A_g$ (5) soft mode is an excellent signature for determining the magnitude of the octahedral rotation in orthorhombic $R$CrO$_3$ and thus for its deviation from the cubic structure. The $A_g$ (5) is preferred to the $A_g$ (3) on the basis of its easy identification throughout the whole series and because the $A_g$ (3) mode is for large $r_R$ likely affected by mode repulsion and intermixing with the $A_g$ (2) mode, as discussed in section III.A.5. This observation in orthochromites, thus in a further family to orthomanganite and orthoscandate perovskites, adds further support to the earlier proposition [26, 55] that a general relationship between the angle of octahedral tilting and the frequency of the rotational soft modes is viable in a vast number of perovskites and thus suggests a general soft-mode type relation of the type $\omega^2 = B(r_{R,c} - r_R) \sim \theta^2$.

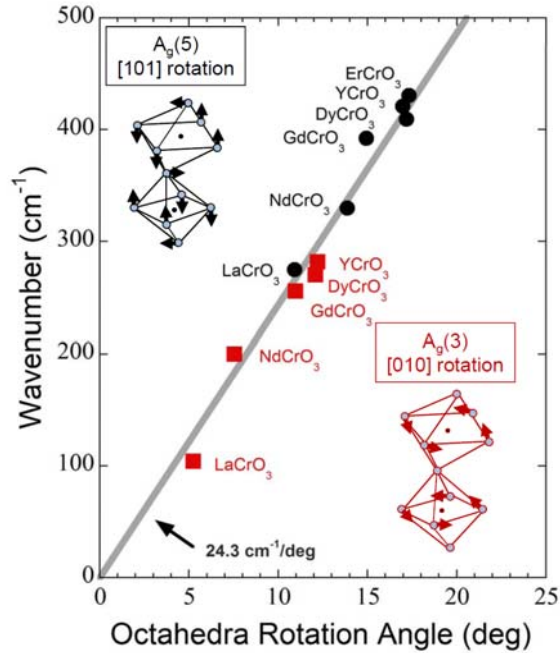

**Figure 7**
Phonon wavenumbers as a function of octahedra tilt angles for $A_g(3)$ [red squares] and $A_g(5)$ [black circles] modes of those $R$CrO$_3$ for which the tilt angle can be calculated from available atomic positions. Raman band positions for NdCrO$_3$ [38] and ErCrO$_3$ [37] are from literature.

# IV. CONCLUSION

We have presented a Raman scattering investigation of a series of nine $R$CrO$_3$ ($R$ = Y, La, Pr, Sm, Gd , Dy, Ho ,Yb, Lu) powder samples. A symmetry assignment of the observed modes has been proposed on the basis of earlier published single crystal data for some orthochromites and by taking advantage of the continuous changes in the Raman spectra across the whole $R$CrO$_3$ series. This careful assignment has allowed verifying the reported literature assignments, assigning the modes for the not yet reported samples, investigating the complex crossing of several modes and, in particular, relating several modes to the structural distortion of the orthochromites. We have shown that the frequency of the $A_g$ (5) correlates in a soft mode fashion with the magnitude of the octahedral rotation in $R$CrO$_3$ and can thus be used to follow the distortion of the structure.

The room-temperature Raman spectra and the associated phonon mode assignment provide reference data for structural investigation of the whole series of $R$CrO$_3$ chromites, including the investigation of strain (via phonon shifts) in thin films. Similarly to previous systematic work on manganites [25] it will be interesting to investigate the $A$-cation dependent spin-phonon coupling in orthochromites via Raman scattering through the magnetic phase transition. We also note that recent *ab-initio* calculations of phonon properties have shown to be a great source of understanding for structural and physical properties of perovskites-type oxides, our study provides benchmark data for such calculations on orthochromites, similarly to recent work in the field [52, 58-59].

**Acknowledgements**

JK is grateful for a visiting fellowship from the Institute of Advanced Study (IAS) Warwick and acknowledges financial support from the department of Physics of the University of Warwick during his sabbatical stay at Warwick. Some of the equipment used at the University of Warwick was obtained through the Science City Advanced Materials project "Creating and characterizing Next Generation Advanced Materials". O. Chaix, Grenoble (F), is acknowledged for a careful reading of the manuscript.